\documentclass{article}

\usepackage{arxiv}

\usepackage[utf8]{inputenc} 
\usepackage[T1]{fontenc}    
\usepackage{hyperref}       
\usepackage{url}            
\usepackage{booktabs}       
\usepackage{amsfonts}       
\usepackage{nicefrac}       
\usepackage{microtype}      
\usepackage{lipsum}		
\usepackage{graphicx}
\usepackage{doi}
\usepackage{makecell}
\usepackage{longtable}

\title{EM and XRM Connectomics Imaging and Experimental Metadata Standards}

\author{ 
  Miguel E. Wimbish, Nicole K. Guittari, Victoria A. Rose, Jorge L. Rivera Jr, Patricia K. Rivlin,\\ 
  \textbf{Mark A. Hinton, Jordan K. Matelsky, Nicole E. Stock, Brock A. Wester,} \\ \textbf{Erik C. Johnson$^{a}$, William R. Gray-Roncal$^{b}$} \\
  Research and Exploratory Development Department, \\
  Johns Hopkins University Applied Physics Laboratory, Laurel, MD, USA \\
  \texttt{$^a$erik.c.johnson@jhuapl.edu; $^b$william.gray.roncal@jhuapl.edu} \\
}

\renewcommand{\shorttitle}{BENCHMARK Standards v1.1}

\hypersetup{
pdftitle={BENCHMARK Metadata Standards v1.1},
pdfsubject={q-bio.NC, q-bio.QM},
pdfauthor={Miguel E. Wimbish, Nicole K. Guittari, Victoria A. Rose, Jorge L. Rivera Jr, Patricia K. Rivlin, Mark A. Hinton, Jordan K. Matelsky, Nicole E. Stock, Brock A. Wester, Erik C. Johnson, William R. Gray-Roncal},
pdfkeywords={Connectomics, Neuroanatomy, Electron Microscopy, X-ray Microtomography, Data Standards, FAIR Data},
}

\begin{document}
\maketitle

\begin{abstract}
	High resolution volumetric neuroimaging datasets from electron microscopy (EM) and x-ray micro and holographic-nano tomography (XRM/XHN) are being generated at an increasing rate and by a growing number of research teams. These datasets are derived from an increasing number of species, in an increasing number of brain regions, and with an increasing number of techniques. Each of these large-scale datasets, often surpassing petascale levels, is typically accompanied by a unique and varied set of metadata. These datasets can be used to derive connectomes, or neuron-synapse level connectivity diagrams, to investigate the fundamental organization of neural circuitry, neuronal development, and neurodegenerative disease. Standardization is essential to facilitate comparative connectomics analysis and enhance data utilization. Although the neuroinformatics community has successfully established and adopted data standards for many modalities, this effort has not yet encompassed EM and XRM/ XHN connectomics data. This lack of standardization isolates these datasets, hindering their integration and comparison with other research performed in the field. Towards this end, our team formed a working group consisting of community stakeholders to develop Image and Experimental Metadata Standards for EM and XRM/XHN data to ensure the scientific impact and further motivate the generation and sharing of these data. This document addresses version 1.1 of these standards, aiming to support metadata services and future software designs for community collaboration. Standards for derived annotations are described in a companion document. Standards definitions are available on a community github page. We hope these standards will enable comparative analysis, improve interoperability between connectomics software tools, and continue to be refined and improved by the neuroinformatics community.
\end{abstract}


\paragraph{BENCHMARK Working Group Participants} Abhishek Bhardwaj, Alyssa Wilson, Ben Dichter, Bill Katz, Chris Broz, Danny Xenes, Hannah Gooden, Josh Morgan, Kabilar Gunalan, Karl Friedrichsen, Kedar Narayan, Kushal Bakshi, Liam McCoy, Mark Hinton, Marta Costa, Norman Rzepka, Nuno da Costa, Oliver Ruebel, Paul Fahey, Sahil Loomba, Sandy Hider, Scott Daniel, Stephan Gerhard, Tim Fawcett, Yaroslav Halchenko, Ming Zhan.

\section{Introduction}
Neuroimaging datasets are rapidly evolving in scale, which may unlock critical insights into neuronal circuit organization and neurodegenerative disease including connectopathies. Supported by large-scale investments, such as the from the National Institutes of Health (NIH) Brain Research Through Advancing Innovative Neurotechnologies (BRAIN) Initiative \cite{jorgenson2015brain}, this growth in scale of neuroimaging is not a niche development as a growing number of research teams and multi-institution teams are generating groundbreaking datasets. 

In particular, there have been major advances in EM and XRM/XNH connectomics dataset generation, which enable the creation of increasingly large connectomes (wiring diagrams) with individual neuron and synapse resolution. Recent efforts have included large scale volumes from multiple species and brain regions, with increasing spatial extent. This includes imaging of C. elegans \cite{white1986structure,witvliet2021connectomes}, Mus musculus cortex \cite{kasthuri2015saturated,bock2011network,dorkenwald2022binary,schneider2020chandelier,microns2021functional}, Drosophila \cite{takemura2013visual,xu2020connectome,dorkenwald2023neuronal}, and Homo sapiens \cite{shapson2021connectomic}. Individual image volumes are reaching the scale of hundreds of terabytes or even petabytes of data. New datasets include co-registration to additional modalities \cite{microns2021functional} as well as imaging of multiple individuals at different developmental time points \cite{witvliet2021connectomes}. In addition to these emerging datasets, there is continued development in imaging techniques, processing algorithms, data storage solutions, and analysis techniques \cite{hider2022brain,dorkenwald2023cave,dorkenwald2023neuronal,turner2022reconstruction,braincircuits,katz2019dvid}. This proliferation of datasets and techniques hold tremendous promise large-scale, cross-dataset discovery, but also presents considerable challenges. 

This growth in neuroimaging is driving the need for sophisticated data management and metadata creation strategies to support the production and dissemination of complex datasets \cite{wilkinson2016fair,johnson2023maturity}. As a result, the field is seeing a significant increase in collaborative neuroinformatics efforts, that are aimed at effectively storing, processing, and sharing vast and intricate datasets. These platforms include, DANDI \cite{rubel2022neurodata}, BrainLife \cite{hayashi2023brainlife}, BossDB \cite{hider2022brain}, BrainCircuits.io \cite{braincircuits}, OpenNeuro \cite{markiewicz2021openneuro}, NEMAR \cite{delorme2022nemar}, BIL \cite{benninger2020cyberinfrastructure}, NEMO \cite{ament2023neuroscience}, and DABI \cite{duncan2023data}. These large-scale neuroinformatics efforts span a range of modalities, including neurophysiology, genomics, transcriptomics, and functional and structural neuroimaging. These datasets are most impactful when they can be accessed and integrated across datasets and even modalities. This has been recently demonstrated by the BRAIN Initiative Cell Consensus Network atlas \cite{network2021multimodal} data ecosystem \cite{hawrylycz2023guide}, which fused data from across archives to create a comprehensive understanding of cellular organization in mammalian cortex. This collaboration among scientific consortia demonstrates the real-world benefits of Findable, Accessible, Interoperable and Reusable (FAIR) principles for data management  \cite{wilkinson2016fair}. These emerging large-scale datasets are challenging to reproduce and reuse, thus a standardized framework is necessary to enable researchers to evaluate data across these datasets \cite{megabrain}. This also applies to emerging EM and XRM/XNH datasets, particularly as these datasets increase in size and complexity.

Given the recent growth in EM and XRM/XNH connectomics datasets, there is now a presenting need for similar neuroinformatics efforts and standardization for connectomics data. To address the needs of the EM and XRM/XNH connectomics community, tools for data storage, visualization and analysis such as BossDB \cite{hider2022brain}, DVID \cite{katz2019dvid}, webKnossos \cite{boergens2017webknossos}, BrainCircuits.io \cite{braincircuits}, cloud-volume \cite{cloudvolume}, Neuroglancer \cite{neuroglancer}, neuPrint \cite{clements2020neu}, and CAVE \cite{dorkenwald2023cave} are being developed and adopted by EM and biology communities \cite{collinson2023volume}. However, there has not yet been a comprehensive effort to standardize EM and XRM/XNH connectomics datasets and provide metadata guidelines for these tools.  

While data standards have been proposed for neurophysiology (Neurodata Without Borders, \cite{rubel2022neurodata}), neuroimaging (BIDS, \cite{gorgolewski2016brain}), and 3D microscopy \cite{ropelewski2022standard}, challenges in data size, data types, and acquisition techniques limit the application of these standards to EM and XRM/XNH connectomics. To overcome these barriers, novel metadata standards must be developed while ensuring harmonization with existing standards for other modalities. To address this, the Big-Data Electron-Microscopy for Novel Community Hypotheses: Measuring And Retrieving Knowledge (BENCHMARK) team formed a working group to create image and experimental metadata standardization recommendations for the EM connectomics community. This document presents the BENCHMARK Image and Experimental Metadata Standards v1.1 for the connectomics community aiming to refine neuroimaging standards to support large-scale programs, archives, and tools. A companion paper describes standards for annotations and data products derived from these data. 

\section{Standards Development Process for Image and Experimental Metadata}

In order to address community concerns about how to structure raw data and metadata for EM connectomics, we formed a community working group. This effort drew from numerous existing neuroscience communities that preserve, manage, and create metadata for different modalities. In the EM Connectomics community, three key areas for standardization were identified as: 
\begin{itemize}
    \item Archive formats for raw image files
    \item Imaging and Experimental Metadata
    \item Annotation data and metadata derived from image data
\end{itemize}
The working group was drawn from active members of the connectomics, EM, and neuroinformatics communities. Representation included students, staff, and investigators. Participants were drawn from the North American and European scientific communities. Universities, non-profit research centers, for-profit companies, and government organizations were represented. The Imaging Working Group contained experimentalists with extensive imaging experience, data scientists, and informatics experts with large scale processing and data storage experience for XRM and EM imaging modalities. Additional meetings were held with individual laboratories as well as small-group discussions on targeted topics.

This document addresses recommendations for 1) Archive formats for raw image files and 2) Imaging and Experimental Metadata standards. Considering the complexity of standards development and community adoption, we envision ongoing review by the community and working group, including open participation and contributions at \url{www.github.com/aplbrain/benchmark-metadata}. We encourage prototype implementations of these standards in EM community tools to enhance accessibility, interoperability, and reusability of data across various archives.

\subsection{EM Connectomics Image and Experimental Metadata Standards Development Process}

The working group met to discuss key guiding principles of standards development in this field, including:
\begin{itemize}
    \item Building on existing file formats and standards efforts when possible
    \item Promoting inter-operability across EM Connectomics archives/tools
    \item Promoting accessibility and findability across modalities
    \item Planning for scale and distributed cloud-based computing
\end{itemize}

During working group meetings, subgroups were formed to discuss image and data file types, linking between datasets, potential multimodal experiments and cross-archive analysis. An emphasis was placed on preserving existing workflows, both within laboratories and existing software tools. Image formats and metadata standards were proposed to address issues during inter-operation (e.g. data ingest), data dissemination, and secondary data analysis. The working group recommended these standards be implemented in software tools for data archives and processing software to enable inter-operability. After each working group meeting, recommendations were aggregated by the core standards team. Draft documents are available at \url{https://github.com/aplbrain/benchmark-metadata} for continued community refinement. 

\subsection{Harmonization Efforts with Existing Standards and Datasets}

\begin{figure}
	\centering
	\includegraphics[width=0.8\linewidth]{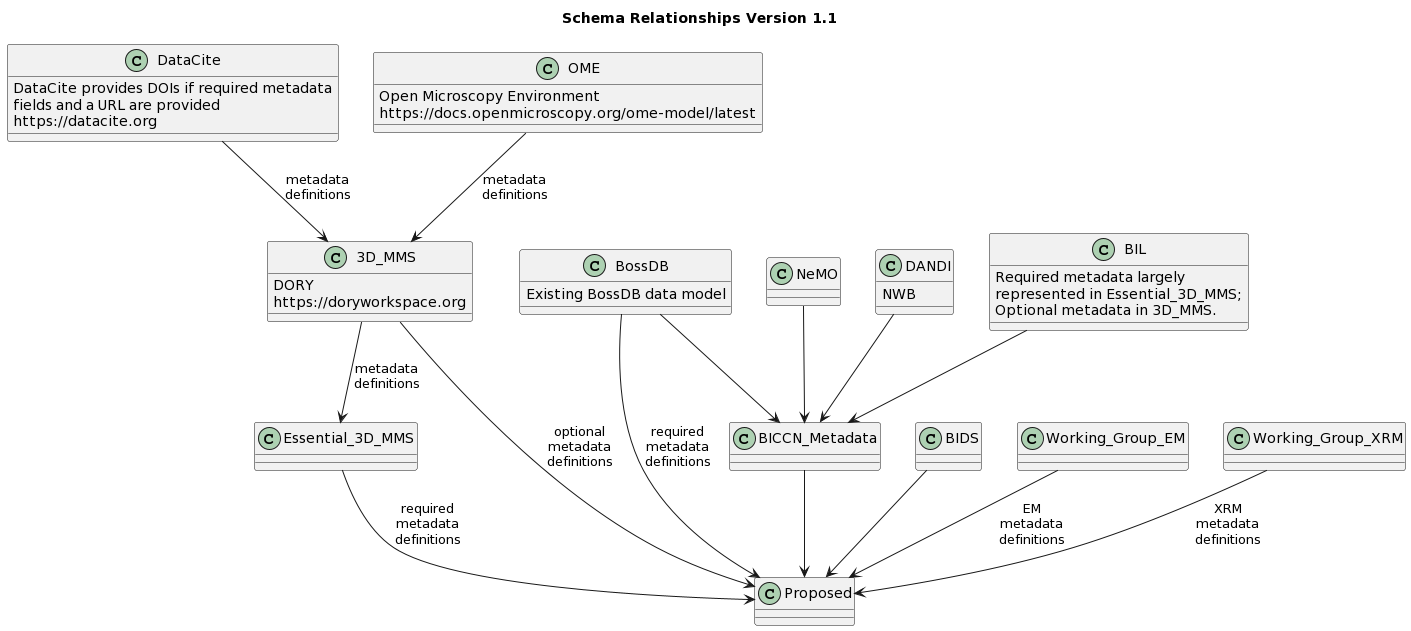}
	\caption{Overview of existing datasets, archives, and standards reviewed during the development process. This approach was critical to harmonization of the proposed standards with existing approaches to maximize reuse and promote interoperability.}
	\label{fig:fig1}
\end{figure}

In addition to working group meetings, extensive review of several existing, related standards and datasets was conducted (Fig \ref{fig:fig1}). This was intended to leverage existing standards, harmonize efforts across archives, and promote a pathway to sustainability, possibly as an extension to an existing standards effort. Continued refinement of these draft standards will focus on harmonization with other metadata specifications. The reviewed standards included:
\begin{itemize}
    \item Metadata services for the archives NEMO \cite{ament2023neuroscience}, DANDI \cite{rubel2022neurodata}, BossDB \cite{hider2022brain}, DViD \cite{katz2019dvid}, webKnossos \cite{boergens2017webknossos} and BIL \cite{benninger2020cyberinfrastructure}
    \item The standard for 3D Microscopy \cite{ropelewski2022standard}
    \item Neurodata Without Borders Standard \cite{rubel2022neurodata}
    \item Brain Imaging Data Structure (BIDS) Standard \cite{gorgolewski2016brain}
    \item Existing standards of the Volume EM community \cite{iudin2023empiar}
\end{itemize}

When possible, metadata fields utilize common terminology with the field. When conflicts exist, a harmonization mapping will be maintained by the BENCHMARK standards community. In addition, 38 extant datasets in the BossDB archive were reviewed. From these, it was determined what project metadata has already been generated. Moreover, a preliminary standardization was created for each dataset. This formed the basis of a prototype implementation of the BENCHMARK Standard (with programmatic RESTful API) at \url{metadata.bossdb.org}. 

Below, we outline v1.1 of the BENCHMARK Image and Experimental Metadata Standards. We anticipate the community will continue to refine and develop this in the future. The metadata is developed to include information on raw image formats, experimental metadata, coordinate frames, image quality, and co-registration to datasets. The proposed metadata captures key aspects of the project: the title, contributors, species, date that the dataset was created, and other information that will be discussed throughout the paper. It also documents the format and structure of the metadata.

\section{BENCHMARK EM Connectomics Standard v1.1}
Based on Working Group input and meetings, we have formulated v1.1 BENCHMARK standards for Image Data and Experimental Metadata. We encourage adoption of this framework by the community, and continued revision to support emerging efforts, needs, and new technologies. These standards are recommended primarily for data dissemination to create FAIR datasets \cite{wilkinson2016fair} (for example on archives, consortia portals, or project websites), rather than prescribing a particular approach to data manipulation within an experimental pipeline. Working versions of the standard are maintained at \url{https://github.com/aplbrain/benchmark-metadata}.

\subsection{Raw Image Formats}
EM and XRM connectomics datasets consist of high-resolution two-dimensional or three-dimensional neuroimaging datasets. Two-dimensional datasets are typically aligned to create a contiguous three-dimensional volume, which is the typical type of data volume stored in online archives. The field of connectomics therefore overlaps heavily with other three-dimensional neuroimaging domains but has particular aspects of data scale, unique proofreading workflows, and secondary data products which require consideration. We aim to leverage existing standards as much as possible, and aim to promote cross-archive and project compatibility. We recognize, however, that individual laboratories and software projects have significant investments in tools and formats and may need to maintain these approaches. However, adhering to the recommended standards on publication/export of the data products can greatly enhance interoperability and reuse in the community and further motivate generation and sharing of new datasets.

\subsubsection{General Desirable Properties of Image and Data Formats}
A subgroup convened to identify key features for data formats to store three-dimensional neuroimaging datasets. Key factors were to create formats suitable for secondary analysis, and export between tools/archives. Key considerations identified were: 
\begin{itemize}
    \item Common, non-evolving formats
    \item Ability to create wrappers around existing tools/ecosystems
    \item Linkages between imagery data and other data types
    \item Fundamental concepts of data sharding \cite{datasharding} should be followed
    \item Consider both the individual laboratory and the needs of a community user
    \item Versioned releases likely sufficient for most secondary analysis
\end{itemize}

\subsubsection{Image Formatting Options}
Promising existing formats were identified which have gained traction within the neuroimaging and biomedical imaging communities. It is recognized existing tools may have and may need to maintain internal data formats, but standard formats are required for interoperability. Version 1.1 of the BENCHMARK standards requires imaging data be stored in one of the following formats:
\begin{itemize}
    \item OME-Zarr (version 2 of the Zarr specification) - to be updated to version 3 of the Zarr specification when appropriate \cite{moore2023ome}
    \item Cloud-volume \cite{cloudvolume}
    \item Neuroglancer precomputed \cite{neuroglancer}
    \item N5 \cite{n5}
\end{itemize}

These formats are broadly supported in the community, and tools such as TensorStore \cite{tensorstore} can be utilized to provide a uniform API for reading and writing multiple array formats, including OME-Zarr, N5, and Neuroglancer precomputed. If a tool/archive uses an internal format, one or more of the above should be supported for compliance with the BENCHMARK v1.1 standard. Key issues to address moving forward include tracking provenance of data files, with immutable logs, versioned releases, and tracking of changes. Additional considerations include label mapping from data files to annotation IDs, an ongoing area of coordination between the raw image formats and annotation metadata formats. Additionally, support for lossless and lossy compression needs to be standardized. 

\subsection{Experimental Metadata for Projects}
\begin{figure}
	\centering
	\includegraphics[width=0.8\linewidth]{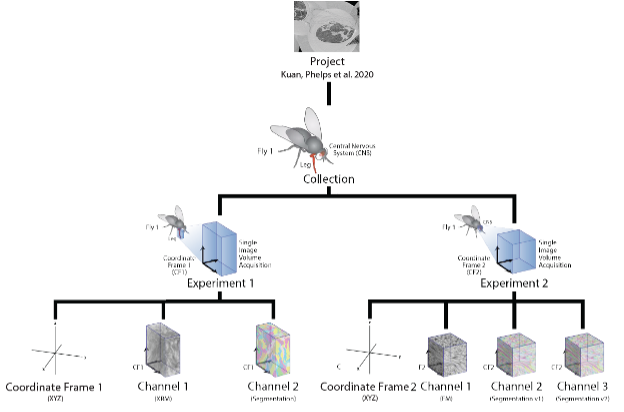}
	\caption{Simplified data model for storage of data, derived from a project hosted at BossDB (\url{https://bossdb.org/project/kuan_phelps2020}). This uses an example dataset including samples from Drosophila in multiple coordinate frames \cite{kuan2020dense} The full metadata structure can be seen at \url{https://github.com/aplbrain/benchmark-metadata} }
	\label{fig:fig2}
\end{figure}

\begin{figure}
	\centering
	\includegraphics[width=0.8\linewidth]{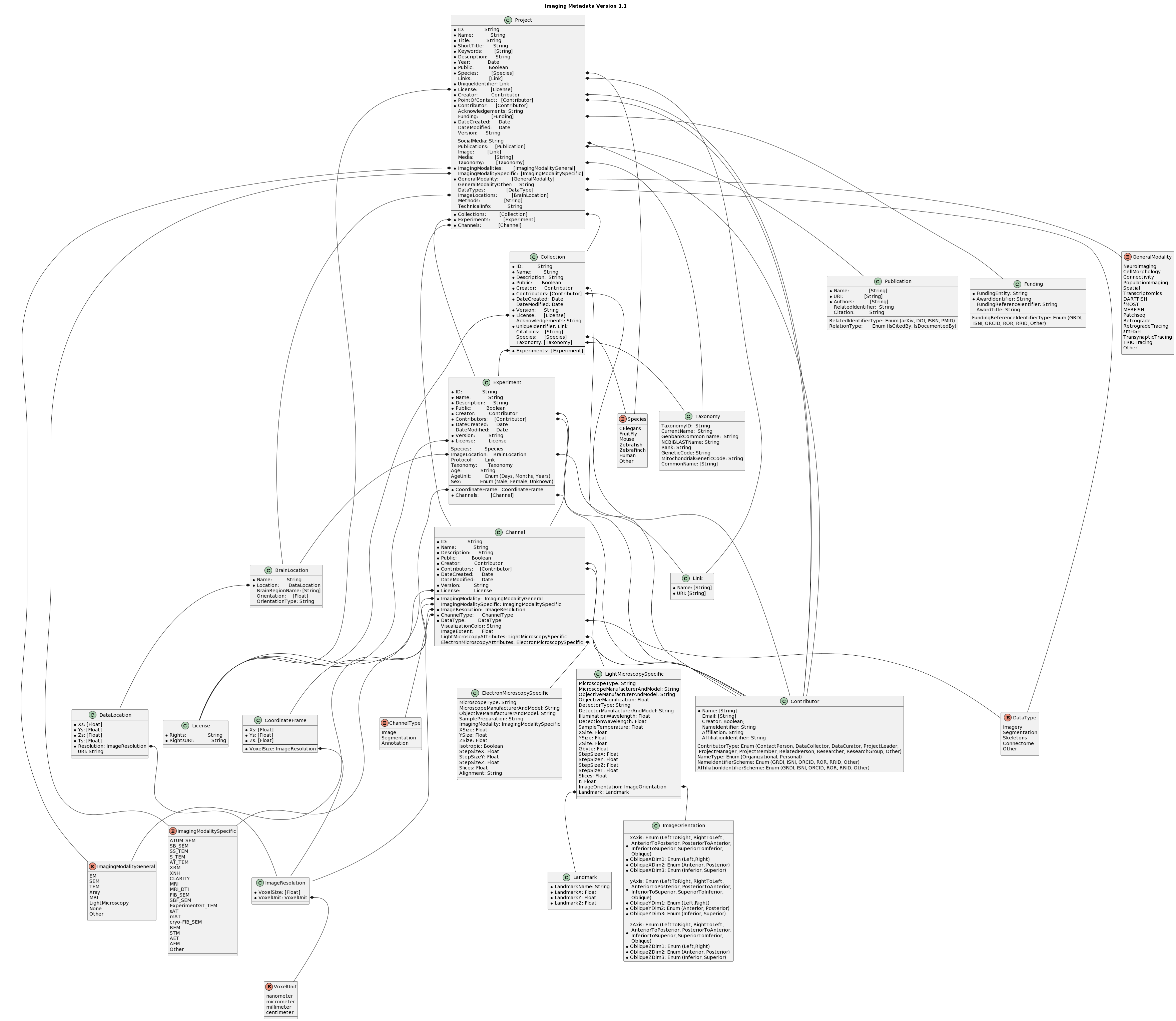}
	\caption{Overview of existing datasets, archives, and standards reviewed during the development process. This approach was critical to harmonization of the proposed standards with existing approaches to maximize reuse and promote interoperability.}
	\label{fig:fig3}
\end{figure}

Fields documented in imaging and metadata standard have 18 classes that are used to denote metadata for a given project. This includes the structure of data channels and experiments within a project (Fig. \ref{fig:fig2}). The metadata fields allow values consisting of strings, integers, enumerations, floats or Booleans, and are essential elements to help define the projects fully. The classes provide fields that gives the field name, description of the field name, and allowed values. The metadata encompasses titles of projects, names of contributors, creators, URIs, and more (Table \ref{tab:table1}). This information can be vital to have when it comes to type of searches and queries, such as project types, specimens’ ages, sex, organizations involved with a particular project, and more. The full relationship of all classes, fields, and enumerations are shown in Fig. \ref{fig:fig3}. 

\subsubsection{Project}
The projects class contains 36 fields that give a high-level description of a dataset. These high-level descriptions include the title, short title, description, year, the DOI (Digital Object Identifier), and more (Table \ref{tab:table1}). The project entity is the highest-level object which gives an overview of the data available in a dataset. This enables users to reference a list of projects with a thorough description and query key project metadata. String identifiers allow for easy search, and embedded identifiers ensure data satisfies FAIR principles. Boolean values represent if a project is public or has a publication, and each project links to related papers and indicates any modification to the project. The project contains links to the collections which make up the project. 

\subsubsection{Collection}
The collection class consists of 18 fields used to associate a set of experiments within a project. A project may contain multiple collections. Collections may associate a set of experiments from a single topic or from a single publication. These may have separate creators, identifiers, and licenses from a project, so long as they are a subset of the overall project creators, identifiers, and licenses. A collection may contain experiments with different coordinate frames (e.g. different brain regions).

\subsubsection{Experiment}
The experiments class consists of 20 fields defining a set of data which share a common coordinate frame. An experiment consists of one or more Channels associated with the coordinate frame. An experiment may have separate creators, identifiers, and licenses from a collection, as long as they are a subset of the overall collection creators, identifiers, and licenses. It also includes a field for an identifier for an experimental protocol. An experiment may contain channels with different data types, such as raw imagery and segmentations. 

\subsubsection{Channel}
The channel class consists of 19 fields defining a three-dimensional volume of imaging data, with a defined coordinate frame. The data may consist of different types, such as unsigned integer or float, but a given channel may only have one type of data. A channel has a uniform resolution for each of the X, Y, and Z dimensions (which may differ between them). A channel may have creators, licenses, and identifiers which are separate from the experiment, as long as they are a subset of the overall experiment creators, identifiers, and licenses. A channel has one associated coordinate frame (which must be the same as the experiment coordinate frame). 

\subsubsection{CoordinateFrame}
The CoordinateFrame class consists of four entities specifying an extent for the X, Y, and Z axis of a three-dimensional volume. Each dimension is specified as a minimum and maximum float value in a list. The resolution of each voxel (in X, Y and Z) is specified for the CoordinateFrame in the VoxelSize field. This is an ImageResolution type which specifies the x,y, and z resolution along with the relevant unit. It is recommended, but not required, that CoordinateFrame objects share a common reference within a project. Future work will consider mappings to common coordinate frameworks, as they are established in the field. 

\subsection{Experimental Metadata Classes}

\subsubsection{BrainLocation}
The BrainLocation class contains 4 fields to capture the location of a coordinate frame within a larger brain volume. This contains information about the spatial position and orientation of a coordinate frame within a brain. It also contains a string name of the region. If this class exists for an experiment, it is recommended, but not required, that the information is consistent with an established coordinate frame. 

\subsubsection{DataLocation}
The DataLocation class contains 5 fields which capture the spatial extent and origin point of a volume within a coordinate frame. It also captures the image resolution in each dimension.

\subsubsection{LightMicroscopySpecific}
This class is derived from the Standard metadata for 3D microscopy \cite{ropelewski2022standard} and is included to promote compatibility with this standard. Future work will investigate further methods for co-registration and harmonization of these standards. The LightMicroscopySpecific class contains 28 fields to capture the properties of channels and experiments imaged with light microscopy techniques. This includes details on the microscope as well as experimental settings. 

\subsubsection{ElectronMicroscopySpecific}
The ElectronMicroscopySpecific class consists of 14 fields which define metadata related to electron microscopes utilized to generate connectivity datasets. This includes information about the microscope itself as well as imaging techniques and settings. 

\subsubsection{Taxonomy}
The taxonomy class contains 8 fields that provide information to categorize and classify biological samples and species. These classes are Taxonomy ID, current name, GenBank common name, NCBI BLAST name, rank, genetic, mitochondrial genetic code, and common name. Taxonomy information can be found on the National Library of Medicine Taxonomy website \cite{ncbitaxonomy}. 
	
\subsubsection{ImageOrientation}
This class is derived from the Standard metadata for 3D microscopy \cite{ropelewski2022standard}, and is used to capture information related to image orientation angles in LightMicroscopySpecific.

\subsubsection{Landmark}
This class is derived from the Standard metadata for 3D microscopy \cite{ropelewski2022standard}, and is used to capture information related to anatomical landmarks in LightMicroscopySpecific.

\subsubsection{ImageResolution}
This class captures the resolution of the voxels in the volume. This is assumed to be uniform throughout the dataset, but is not assumed to be isotropic. Units for the resolution are also specified.

\subsection{Project Metadata Classes}

\subsubsection{Contributor}
The contributor class consists of 11 fields that provide information about personnel associated with a specific project. This class gives information on the individual or group members who provided skills, knowledge, and resources for a particular project. The standard specifies the role that the individuals played towards the project, and each project is assigned a point of contact. There is also a minimum of one creator for each project. Out of the 11 fields, 6 fields in the contributor class are associated with digital identifiers. These identifiers include ORCID (Open Researcher and Contributor Identifier), RRID (Research Resource Identifiers), ISNI (International Standard Name Identifier), ROR (Research Organization Registry), and a few others (Table \ref{tab:table1}).

\subsubsection{License}
Within the license class, there are 4 fields that give information about the licenses which govern usage of the data for this project. This specifies the terms and agreements that protect the rights of a project. In the license class (Table \ref{tab:table1}), the fields that are required include the URI to the rights, identifier, and corresponding DOI. The project may be associated with multiple licenses, which reference individual datasets within the given project. For example, all public projects on the archive BossDB are currently licensed under the Creative Commons Attribution 4.0 International License \cite{creativecommons}.

\subsubsection{Funding}
The funding class contains 5 fields explaining the organization(s) that financially support a given project. The funding class will provide information regarding what corporation or grant provided resources to support the project. These resources can come from various sources (e.g., government, individual donors, etc.), and the fields that are included in the funders class are funding entity, award identifier, funding reference identifiers, funding resources identifiers type, and award title.

\subsubsection{Publication}
In the publication class, there are 8 fields capturing information about a dataset, papers, and where publications can be found. This allows for tracking the impact of the work with a particular dataset. The fields that are included in the publication class are URI, authors, related identifiers, PubMed Central Identification (PMCID), relation type, and citation. This class is critical to give creditability and validity to the team which generated the dataset.

\subsubsection{Link}
Within the link class, there are two fields. These field names are name and Uniform Resource Identifier (URI), and are used to capture external references and resources.

\subsection{Enumerations}
Several enumerations are defined for restricted fields in the standard. These include
\begin{itemize}
    \item Species – common names of species used in a project, intended for human-interpretable search
    \item DataType – Data products included in a channel, such as segmentations or raw data
    \item ChannelType – Classifies a channel as raw data, a segmentation, or an annotation channel
    \item GeneralModality – Defines the general experimental modality of the dataset
    \item ImagingModalityGeneral – Defines the general imaging techniques used in the dataset
    \item ImagingModalitySpecific – Defines specific imaging types used in the dataset (e.g. FIB-SEM)
    \item VoxelUnit – Defines resolution units
\end{itemize}

\section{Discussion}
The BENCHMARK Image and Experimental Metadata Standards represent a preliminary consensus, drawing from our working group and existing standards efforts, to establish a robust framework for EM and XRM/XNH connectomics. The unique challenges inherent to EM data, such as immense size and diverse acquisition methods, demand specialized standards to ensure data integrity and interoperability. These standards address raw image formats, experimental metadata, coordinate frames, and publication details and underscores the commitment to comprehensive data management. These standards not only facilitate adherence to the FAIR principles but also set the stage for more streamlined collaboration and data sharing within the EM connectomics community.

The proposed standards build on and incorporate existing standards efforts in the community, including existing BRAIN Initiative Archives \cite{rubel2022neurodata, hider2022brain, benninger2020cyberinfrastructure, ament2023neuroscience}, deployed solutions from the EM connectomics community \cite{katz2019dvid,cloudvolume,neuroglancer,clements2020neu,dorkenwald2023cave,braincircuits}, and existing standards efforts \cite{rubel2022neurodata,ropelewski2022standard, gorgolewski2016brain}. The incorporation of existing file formats and standards in the BENCHMARK recommendations demonstrates a pragmatic approach, acknowledging the importance of building upon established
practices. This strategy not only promotes interoperability among various EM Connectomics archives and tools but also facilitate the seamless integration of these standards into the existing research infrastructure. By prioritizing compatibility, the BENCHMARK standards aim to reduce barriers to adoption, ensuring a smooth transition for researchers while enhancing the overall accessibility and utility of EM connectomics data. This will be particularly critical to enabling multi-modal analysis \cite{hawrylycz2023guide} and secondary comparative connectomics analysis \cite{barsotti2021neural}.

In the context of data scalability, the BENCHMARK standards exhibit a forward-looking perspective by emphasizing compatibility with cloud-based computing. For example, a preliminary prototype implementation is available (\url{metadata.bossdb.org}) which allows programmatic access of project metadata. This service is directly usable by individuals, but also allows for cross-archive indexing, creating real-time dashboards, and contributing to scientific gateways of large consortium projects. 

We encourage continued evolution of these standards with input from the community, developers of tools, and emerging scientific consortia. Further work will be required to fully capture novel and unique experimental and imaging approaches, track software tools and methods used to process the data, and support emerging comparative experiments. Open community comment, discussion, and contributions are encouraged at the BENCHMARK standards community repository (\url{www.github.com/aplbrain/benchmark-metadata}). Future working group meetings will continue to oversee governance and development of new versions.  

While evolution of these approaches is expected, this version of the BENCHMARK Image and Experimental Metadata Standards aims to provide a consistent set of image and experimental metadata information for the unique and emerging fields of EM and XRM connectomics. By addressing key aspects of data management and compatibility these standards contribute to the advancement of secondary analysis as well as cross-modality analysis. In particular, we hope these efforts will support emerging large-scale scientific efforts such as the NIH BRAIN CONNECTS program.  

\section*{Acknowledgments}
We would like to acknowledge and thank our BENCHMARK working group participants, listed above, for their invaluable contributions and expertise. Their contributions formed the core of the suggested standards, and we are honored to collaborate with these individuals. We would like to thank the BRAIN Informatics program archives, the BICCN consortium, the NWB standards group, the BIDS standards group, and the 3D-MMS group for developing and maintaining their valuable metadata standards. This work was supported in part by National Institutes of Health (NIH) grants R24MH114799, R24MH114785, and R01MH126684. The content is solely the responsibility of the authors and does not necessarily represent the official views of the National Institutes of Health.

\begin{longtable}{llll}

		\toprule
		Field Name & Description & Data Type & Required \\
            \midrule
            \multicolumn{4}{l}{Contributor}             \\
		\midrule
		Name & \makecell[l]{Individuals, organizations, or entities \\ who contributed to or are responsible \\ for a project} & [String] & Yes \\
		Email & Contact email & [String] & No \\
            Creator & Identifies contributor who created a dataset & Boolean & No \\
            ContributorType & \makecell[l]{Categorization of the role of the contributor. \\ Recommended: ProjectLeader, \\ ResearchGroup} & \makecell[l]{Enum (ContactPerson, \\ DataCollector, \\ DataCurator, \\ ProjectLeader, \\ ProjectManager, \\ ProjectMember, \\ RelatedPerson, \\ Researcher, \\ ResearchGroup, \\ Other)}  & No \\
            NameType & Type of contributor & \makecell[l]{Enum (Organizational, \\
Personal)} & No \\
            NameIdentifier & \makecell[l]{Identifier of individual or\\ entity that created contribution} & String & No \\
            \makecell[l]{NameIdentifier \\ Scheme} & \makecell[l]{Identifying scheme \\ used in NameIdentifier} & \makecell[l]{Enum (GRDI, ISNI, \\ ORCID, ROR, RRID, \\ Other)} &	No \\
            Affiliation & \makecell[l]{Organization associated with \\individual contributor a particular project} & String & No \\
            AffiliationIdentifier & \makecell[l]{Affiliation Identifiers are \\ unique values assigned to affiliation} & String & No \\
            \makecell[l]{AffiliationIdentifier \\ Scheme}	& \makecell[l]{Identifying scheme used \\ in AffiliationIdentifier} & \makecell[l]{Enum (GRDI, ISNI, \\ ORCID, ROR, \\ RRID, Other)} & No \\
            \midrule
            \multicolumn{4}{l}{License}             \\
            \midrule
            Rights & \makecell[l]{License which defines \\ usage of the data} &	String & Yes\\
            RightsURI & \makecell[l]{The rightsURI is a digital \\  resource providing information \\ about the license of \\ a project} & String & Yes \\
		\midrule
            \multicolumn{4}{l}{Funding}             \\
            \midrule
            FundingEntity & \makecell[l]{The individual or organization \\ who are providing financial support \\ to a project} & 	String & Yes \\ 
            AwardIdentifier & \makecell[l]{Award identifier provides an \\ award or grant number or name} & String & Yes \\
            \makecell[l]{FundingReference \\ Identifier} & \makecell[l]{Identifier for the funding source} & String & No \\
            \makecell[l]{FundingReference \\ IdentifierType} & \makecell[l]{Type of funding \\ source identifier} & \makecell[l]{Enum (GRDI, \\ ISNI, ORCID,\\ ROR, RRID,\\ Other)} & No \\
            AwardTitle & \makecell[l]{The name of the award or grant \\  from funding entity} & String & No \\
            \midrule
            \multicolumn{4}{l}{Projects}             \\
            \midrule
            Title & Title for a specific project & String & Yes \\
            ShortTitle & \makecell[l]{Display title of less \\ than 100 characters} & String & Yes \\
            ID & \makecell[l]{Provides short, unique \\ identifier for project} & String & Yes \\
            Keywords & Keyword descriptors for the project & [String] & Yes\\
            Description & \makecell[l]{Provides information of species \\type, image modality, and brief \\ definition of the project} & String & Yes \\
            Public & \makecell[l]{Indicates whether the dataset \\ is public (true) or restricted (false)} & Boolean & Yes \\
            Year & Year project was created & Date & Yes \\
            Publications & Publications linked to project & [Publication] & No \\
            Links & External links associated with project & [Link] & No \\
            License & \makecell[l]{One or more licenses associated \\ with the use of this project} & [License] & Yes \\
            Creator & Individual who created this project & Contributor & Yes \\
            PointOfContact & \makecell[l]{Point of contact for this \\ project, may be multiple people} & [Contributors] & Yes \\
            Contributor & \makecell[l]{Persons who contributed to \\ creation of this project} & [Contributor] & Yes \\
            Acknowledgements & Acknowledgements statement & String & No \\
            Funding & Information about funding agencies & [Funding] & No \\
            Species & Organism type & [Species] & Yes \\
            Taxonomy & Taxonomy of organisms in project & [Taxonomy] & No \\
            DateCreated & \makecell[l]{Gives the date for when the\\  project was created} & Date & Yes \\
            Channels & \makecell[l]{Channels associated with this \\ project. Must be at least one} & [Channel] & Yes \\
            Experiments & \makecell[l]{Experiments associated with \\ this project. \\ Must be at least one} & [Experiments] & Yes \\
            Collections & \makecell[l]{Collections associated with \\ this project. \\ Must be at least one} & [Collection] & Yes \\
            Media & Associated media information or links & [String] & No\\
            DataTypes & \makecell[l]{Types of data contained in \\experiments associated with this project}	& [DataType] & No \\
            Image & \makecell[l]{Files or links of images associated \\ with this project, for display}	& [Link] & No \\
            TechnicalInfo & Additional technical information & String & No \\
            Methods & \makecell[l]{Identifiers linking to description \\ of methods (paper or \\ protocols.io, for instance)} & [String] & No \\
            SocialMedia & \makecell[l]{Website, social media account, \\ or description associated with project} & String & No \\
            GeneralModality & \makecell[l]{Gives list of general techniques \\ and approaches associated with this\\ dataset. This may indicate \\ additional modalities associated with \\ the project, beyond microscopy data} & [GeneralModality] & Yes\\
            GeneralModalityOther & \makecell[l]{Specification of other modalities \\ not described in standard} & String & No \\
            ImagingModalities & \makecell[l]{Gives a list of modalities \\ contained in this project} & \makecell[l]{[ImagingModality\\General]} & Yes \\
            \makecell[l]{ImagingModality\\Specific} & Specific imaging types used in this project & \makecell[l]{[ImagingModality\\Specific]} & No\\
            ImageLocations & \makecell[l]{Locations of experimental volumes \\ within the brain} & [BrainLocation] & No\\
            Version & Current version of project & String & Yes \\
            UniqueIdentifier & \makecell[l]{Unique identifier for dataset \\ (DOI preferred)} & Link & No\\
            DateModified & \makecell[l]{Gives the date for when the \\project was updated or changed} & Date & No \\
            \midrule
            \multicolumn{4}{l}{Taxonomy}             \\
            \midrule
            TaxonomyID & \makecell[l]{A unique numerical identifier \\ that is assigned \\ to a specific organism \\ in the NCBI Taxonomy database} & Int & Yes \\
            CurrentName & \makecell[l]{The organism type in its \\ scientific form. Ex: \\ Mus musculus, Drosophila, C. elegans} & String & No \\
            GenBankCommonName & \makecell[l]{Provides a simplified and  \\ recognizable name for \\ a specific organism} & Integer & No \\
            NCBIBlastName & \makecell[l]{Blast name provided by NCBI}	& String & No \\
            Rank & \makecell[l]{Categorize and organize organisms \\ based on their evolutionary relationships.\\ Ex. Species, genus, family, etc} & String & No \\
            GeneticCode & \makecell[l]{Genetic information regarding the species} & String & No \\
            \makecell[l]{Mitochondrial\\GeneticCode} & \makecell[l]{Mitochondrial genetic information \\ regarding the species.} & String & No \\
            CommonName & \makecell[l]{The everyday name of a \\
            species, organism, or biological entity} & [String] & No \\
            \midrule
            \multicolumn{4}{l}{CoordinateFrame}             \\
            \midrule
            Xs & \makecell[l]{Numerical value represented \\ horizontally in two-dimensions} & [Float] & Yes \\
            Ys & \makecell[l] {Numerical value represented vertically \\ in two-dimension} & [Float] & Yes \\
            Zs & \makecell[l]{Numerical value representing height \\ in three-dimension} & [Float] & Yes \\
            VoxelSize & \makecell[l]{The resolution of each \\ 3D voxel} & ImageResolution & Yes \\
            \midrule
            \multicolumn{4}{l}{Publication}             \\
            \midrule
            Name & Name of publication & [String] & Yes \\
            URI & \makecell[l]{Uniform Resource Identifier \\ for affiliated web resource} & [String] & Yes \\
            Authors & \makecell[l]{The creators or writers \\ of the work} & [String] & Yes \\
            RelatedIdentifier & Identifier for publications. & String & No \\
            RelatedIdentifierType & Type of publication identifier. & \makecell[l]{Enum (arXiv, DOI, \\ ISBN, PMID, Other)}  & No \\
            RelationType & Relationship to publication. & \makecell[l]{Enum (IsCitedBy, \\ IsDocumentedBy)} & No \\
            Citation & Preferred citation for this publication. & String & No \\
             \midrule
            \multicolumn{4}{l}{Link}             \\
            \midrule
            Name & Provides the name of website or resource & [String] & Yes \\
            URI & \makecell[l]{Uniform Resource Identifier pathway\\ to the website or resource} & [String] & Yes \\
        \bottomrule \\
        \caption{An overview of key metadata classes required for the         BENCHMARK standards. It details various fields such as C    Contributors, License, Funder, Projects, Taxonomy, Coordinate Frame, Publication, and Link, each with specific data types and requirements. This table serves as a reference for understanding the structure of metadata in connectomics, aiding in the organization, retrieval, and analysis of neuroscience data.}
	\label{tab:table1}
\end{longtable}

\begin{longtable}{ll}

		\toprule
		Abbreviation & Description \\
            \midrule
            \multicolumn{2}{l}{Techniques}             \\
		\midrule
            AET & Analytical Electron Microscopy \\
            AFM & Atomic Force Microscopy \\
            AT\_TEM & Array Tomography Transmission Electron Microscope \\
            ATUM\_SEM & Automated Tape Collecting Ultramicrotome Scanning Electron Microscope \\
            CLARITY & \makecell[l]{Clear Lipid-exchanged Acrylamide-hybridized Rigid Imaging / Immunostaining / \\ in situ-hybridization-compatible Tissue hYdrogel} \\
            Cryo\_FIB\_SEM & Cryogenic focused Ion Beam-Scanning Electron Microscope \\
            FIB\_SEM & Focused Ion Beam-Scanning Electron microscope \\
            LEEM & Low-Energy Electron Microscopy \\
            LM & Light Microscopy \\
            PEEM & Photoemission Electron Microscopy \\
            mAT & Magneto Acoustic Tomography \\ 
            MRI & Magnetic Resonance Imaging \\
            REM & Reflection Electron Microscope \\
            sAT & Scanning Acoustic Tomography \\
            SB\_SEM & Serial Block-Face Scanning Electron Microscope \\
            SEM & Scanning Electron Microscope \\
            S\_TEM & Scanning Transmission Electron Microscopy \\
            SS\_TEM & Serial Section Transmission Electron Microscopy \\
            STM & Scanning Tunneling Microscope \\
            TEM & Transmission Electron Microscope \\
            XRM & Xray Microscopy \\
            \midrule
            \multicolumn{2}{l}{Identifiers}             \\
		\midrule
            GRID & Global Research Identifier Database \\
            ISNI & International Standard Name Identifier \\
            ORCID & Open Researcher and Contributor ID \\
            PMCID & PubMed Central Identification \\
            ROR & Research Organization Registry \\
            RRID & Research Resource Identifier \\
            URI & Uniform Resource Identifier \\
            \midrule
            \multicolumn{2}{l}{Other}             \\
		\midrule
            NCBI & National Center for Biotechnology Information \\
            \bottomrule
            \caption{This chart serves as a reference of commonly used acronyms for the fields used in the BENCHMARK standards. }
            \label{tab:table2}
\end{longtable}

\bibliographystyle{ieeetr}
\bibliography{template}  






\end{document}